\documentclass[aps,prl,twocolumn,showpacs,preprintnumbers,amsmath,amssymb,floatfix]{revtex4-1}

\usepackage{graphicx}

\usepackage{dcolumn}
\usepackage{bm}

\newcommand\Rey{\mbox{\textit{Re}}}  

\usepackage[dvipsnames]{xcolor}  
\definecolor{linkcolor}{rgb}{0,0,0.42} 

\usepackage[]{natbib}
\usepackage[
    bookmarksopen=true,
    linktocpage=true,
    urlcolor=linkcolor, 
    citecolor=linkcolor, 
    linkcolor=linkcolor, 
    pdfpagelabels=true,
    plainpages=false,
    colorlinks=true, 
    bookmarks=true,
    pdfview=FitB]{hyperref}

\begin{document}
\title{The chaotic dynamics of large-scale structures in a turbulent wake}
\author{Eliott Varon}
\email[]{eliott.varon@espci.fr}
\author{Yoann Eulalie}
\author{Stephie Edwige}
\author{Philippe Gilotte}
\author{Jean-Luc Aider}
\email[]{jean-luc.aider@espci.fr}
\homepage[]{https://blog.espci.fr/jlaider/}

\affiliation{Laboratoire PMMH, CNRS UMR7636, ESPCI Paris, PSL Research University, 10 rue Vauquelin, Paris, France}
\affiliation{Laboratoire PMMH, CNRS UMR7636, Universit\'e Pierre et Marie Curie, 10 rue Vauquelin, Paris, France}
\affiliation{Plastic Omnium Auto Ext\'erieur Services (POAES), Parc Industriel de la Plaine de l'Ain, Sainte-Julie, France}

\textit{Accepted by Physical Review Fluids. Last version: \today .}

\begin{abstract}
The dynamics of a 3D bimodal turbulent wake downstream a square-back Ahmed body are experimentally studied in a wind-tunnel through high-frequency wall pressure probes mapping the rear of the model and a horizontal 2D velocity field. The barycenters of the pressure distribution over the rear part of the model and the intensity recirculation are found highly correlated. Both described the most energetic large-scale structures dynamics, confirming the relation between the large-scale recirculation bubble and its wall pressure foot-print. Focusing on the pressure, its barycenter trajectory has a stochastic behavior but its low frequencies dynamics exhibit the same characteristics as a weak strange chaotic attractor system, with two well defined attractors. The low frequencies dynamics associated to the large-scale structures are then analyzed. The largest Lyapunov exponent is first estimated, leading to a low positive value characteristic of strange attractors and weak chaotic systems. Afterwards, analyzing the autocorrelation function of the time-series, we compute the correlation dimension, larger than two. The signal is finally transformed and analyzed as a telegraph signal showing that its dynamics correspond to a quasi-random telegraph signal. This is the first demonstration that the low frequencies dynamics of a turbulent 3D wake are not a purely stochastic process but rather a weak chaotic process exhibiting strange attractors. From the flow-control point of view, it also opens the path to more simple closed-loop flow control strategies aiming at the stabilization of the wake and the control of the dynamics of the wake barycenter. 
\end{abstract}

\maketitle

\section{Introduction}

It is well-known that the turbulent wakes downstream 3D bluff-bodies can be very complex, exhibiting large-scale and small-scale coherent structures with strongly intermittent behaviors. Among the various 3D bluff-bodies, one of the most famous is the so-called ``Ahmed body'' which is a model used in automotive aerodynamics to study the wake of a very simplified passenger car \cite{Ahmed1984}. Depending on the geometry of the rear part, the overall structure of the wake changes together with the aerodynamic drag coefficient. One can find a competition between large-scale streamwise longitudinal vortices \cite{Beaudoin2004}, spanwise Kelvin-Helmholtz vortices, recirculation bubbles or toro\"{\i}dal vortices. If the time-averaged velocity fields are relatively simple and well-defined, the instantaneous velocity fields are very complex and exhibits together large and small-scale structures leading to one of the most complex 3D turbulent flows. Recently, it has been shown experimentally \cite{Grandemange2013a} and numerically \cite{Oesth2014} that square-back Ahmed body at high Reynolds numbers exhibits a peculiar behavior with a  bimodal wake, which was first observed in the laminar regime \cite{Grandemange2012a}. Indeed, depending on the geometric parameters (aspect ratio of the bluff-body's cross-section \cite{Grandemange2013b}, underbody flow \cite{Cadot2015}), one can observe a right-left oscillation of the global wake, defining the so-called Reflectional Symmetry Breaking (RSB) modes. 

The large-scale dynamics of another wake, which is 3D turbulent axisymmetric, has been captured by \cite{Rigas2015}, using a deterministic model for the persistent laminar instabilities coupled with a stochastic representation of the turbulent fluctuations. For such dynamics two different time scales are notable: a short one is associated to the vortex shedding process whereas the symmetry breaking are characterized by a long time scale \cite{Grandemange2013a,Rigas2014}. We are interested here in the characterization of the bimodal oscillation of the near wake in the framework of dynamical systems theory and, more precisely, as a chaotic system. 

A classic example of a chaotic system is the Lorenz attractor and corresponding Lorenz system, whose characteristic butterfly shape is famous \cite{Lorenz1963}. The so-called Lorenz system is a simplified ``weather model'' defined by the set of three ordinary differential equations:
\begin{equation}
\begin{cases}
	 \dot{x}=\sigma (y-x) \\
	 \dot{y}=x (R-z)-y \\
	 \dot{z}=xy-\beta z
\end{cases}.
\label{eq:LorenzDef}
\end{equation}
With the correct choice for the three parameters ($\sigma=16$, $R=45.92$ and $\beta=4$), the trajectory plotted in the (x, y, z) space exhibits a chaotic behavior, circling in an apparent random manner between two stable attractors.  

Since the pioneering work of Lorenz, it has been shown that many biological, natural or artificial systems, either at very small or very large scales, follow a chaotic dynamics. The brain wave activity (EEG) \cite{Achermann1994} or the heart rate activity (ECG) \cite{Fojt1998} can exhibit chaotic behaviors. In some cases chaotic excitations can be used to study the response of a mechanical systems. A variation in correlation dimensions can be used as an indicator of a fracture in the overall structure \cite{Yan2013}. Chaotic behavior has been found in trading market time-series \cite{Orlando2016}. One can also find chaotic behavior for large-scale phenomena like earthquakes \cite{Matcharashvili2002}.

In the following, we will first show how the 3D full turbulent wake dynamics can be characterized by the single trajectory of its projected barycenter. After recovering the classic chaotic pattern, we will analyze more thoroughly the inner characteristics of the large-scale dynamical system. We will, in particular, evaluate the family of random process to which it belongs, the largest Lyapunov exponent of the system and the correlation dimension.

\section{Experimental set-up}

\subsection{Ahmed body}

The bluff body is a 0.7 scale of the original Ahmed body ($L=0.731$ m long, $H=0.202$ m high and $W=0.272$ m wide), as described in \cite{Eulalie2014}. The rear part of the model is a square-back geometry with sharp edges.

\subsection{Wind-tunnel}

Experiments are carried out in the PRISME laboratory wind-tunnel (Orl\'eans, France). The model is mounted on a raised floor with a properly profiled leading edge and an adjustable trailing edge to avoid undesired flow separations. The ground clearance is set to $C=5$ cm. In the following, the free-stream velocity is $U_{\infty} =~30$ m.s$^{-1}$, which corresponds to a Reynolds number based on the height of the model $\Rey_H={U_{\infty}H}/{\nu_{air}}=3.9\times10^5$ where $\nu_{air}$ is the kinematic viscosity of the air at ambient temperature. The origin is located on the rear of model ($x=0$), in the vertical symmetry plane ($y=0$) and on the raised floor ($z=0$). Nondimensionalization is applied to distances such as $x^*=x/H$, $y^*=y/H$ and $z^*=z/H$.

\subsection{Sensors}

\begin{figure}[h]
\setlength{\unitlength}{1cm}
\includegraphics[width=\linewidth]{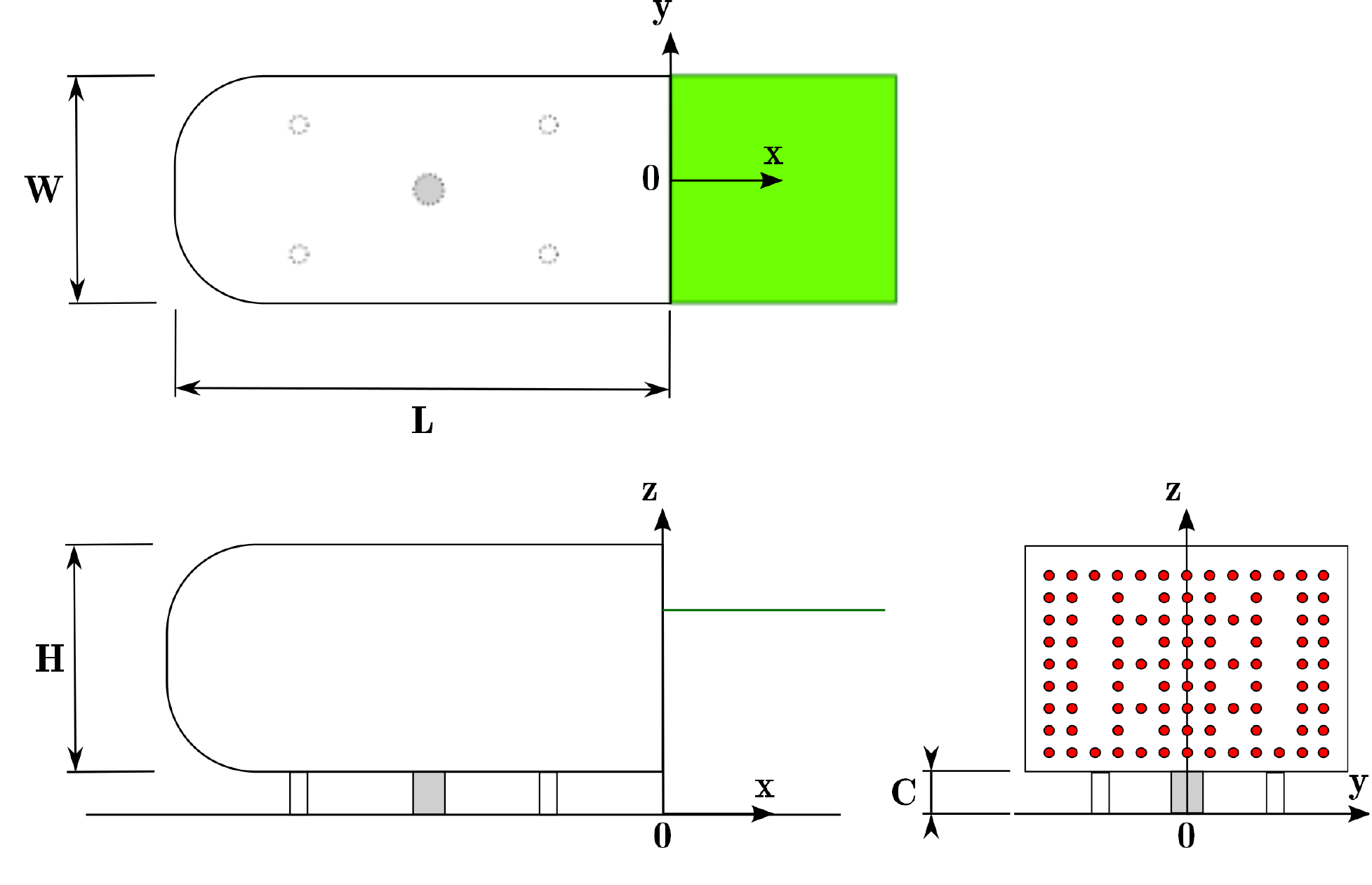}
\caption{Upper view (top figure) and side view (lower left figure) and view from behind (lower right figure) of the Ahmed body. The PIV measurement plane is shown on  top (green rectangle) and lower left (green line) figure. The rear part of the model is mapped with 95 pressure sensors (red circles on lower right figure).  The bluff-body is fixed on the aerodynamic balance through a supplementary leg (grey part on lower figures).}\label{fig:Ahmed body}
\end{figure}

\begin{figure}[h]
\setlength{\unitlength}{1cm}
\includegraphics[width=\linewidth]{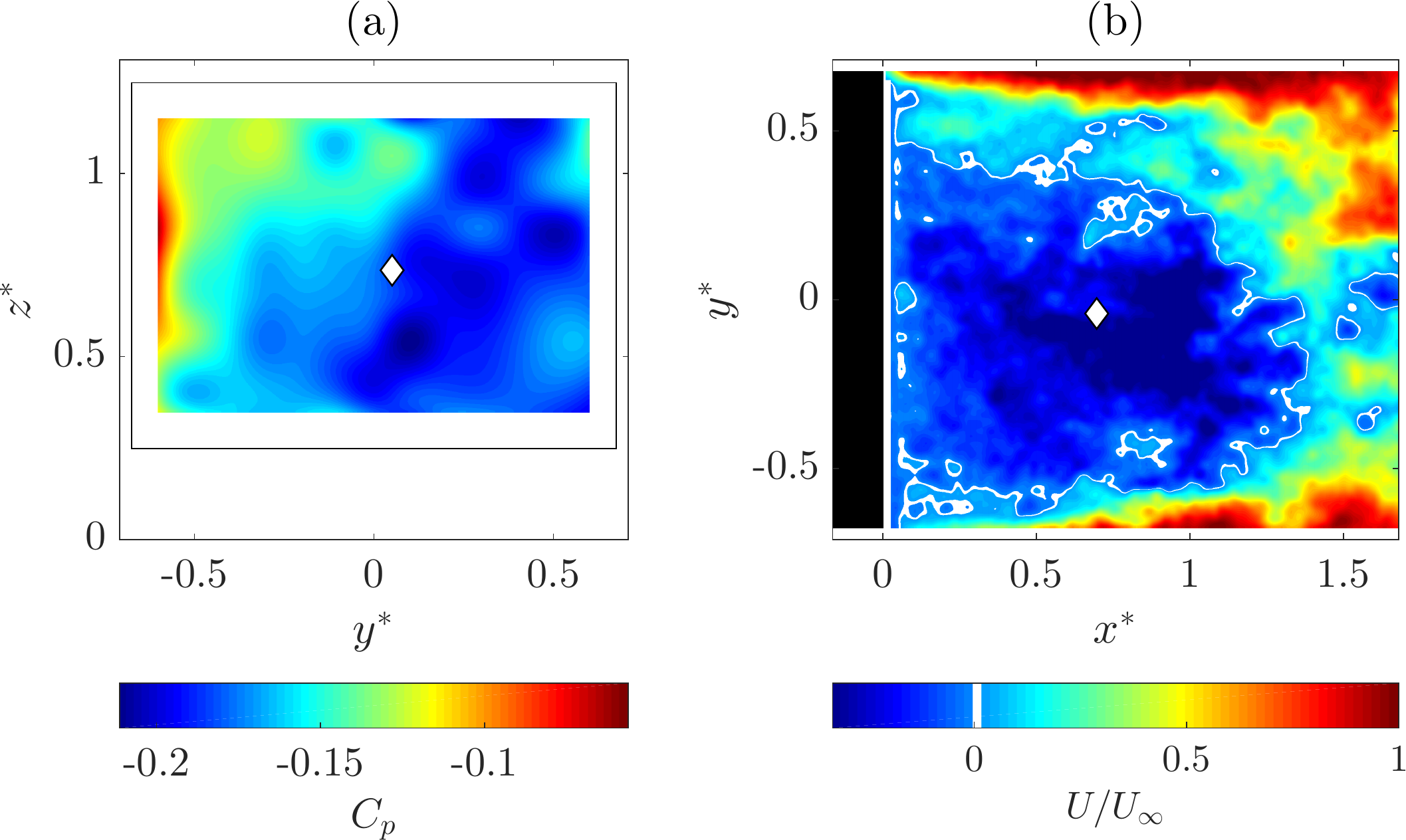}
\caption{Typical (a) instantaneous pressure field over the rear part of the model and (b) instantaneous 2D velocity field in the horizontal plane shown on Fig.~\ref{fig:Ahmed body}. The diamond (white) is the instantaneous barycenter position of respectively (a) the rear pressure $G_p$ and (b) the intensity recirculation $G_{rec}$, which are introduced and dicussed in the following section.}\label{fig:Cp_PIV}
\end{figure}

The wall-pressure over the rear part of the model is studied using a set of 95 pressure vinyls defining an area denoted $S_p$ and covering 70$\%$ of the entire surface $S_r$, as shown on Fig.~\ref{fig:Ahmed body}. Each vinyl is 2 cm away from each of its neighbors and is connected to a 32-channels microDAQ pressure scanner insuring an accuracy of $\pm 17$ Pa located inside the body. The number of samples per acquisition is unfortunately bounded to $N=3\times10^4$, making the sampling frequency for the pressure $f_{P}$ dependent on the acquisition time $T_{P}$: {$f_{P}=N/T_{P}$}. A typical instantaneous pressure field is shown on Fig.~\ref{fig:Cp_PIV}~(a). From these instantaneous pressure fields, a global indicator of the state of the wake can be inferred, as it will be detailed in the following section.

The velocity fields are obtained using a standard PIV (Particle Image Velocimetry) set-up based on a double-frame $14.50$~Hz TSI camera streaming snapshots on a computer and synchronized with a double-cavity pulsed YaG laser. The investigated PIV plane is the near wake horizontal plane at $z^*=1$.

The 2D velocity fields are computed at the frequency $f_{PIV}=4$ Hz using an optical flow algorithm implemented on a GPU. The interrogation window size is $16 \times 16$ pixels and the calculation is based on three iterations for each of the three pyramid reduction levels. One can find more details on this measurement method in \cite{Champagnat2011,Gautier2014,Gautier2015a,Gautier2015b,Pan2015} which rigorously demonstrate its offline accuracy as well as its online efficiency in closed-loop flow control experiments. An example of a 2D instantaneous velocity field is shown on Fig.~\ref{fig:Cp_PIV}~(b). One can see the complexity of the turbulent wake, with large and small-scale strongly fluctuating vortices.

The bluff body is linked to an aerodynamic balance through a cylindrical leg of diameter 32 mm localized at the center of the bottom face. This leg does not modify the reflectional symmetry of the squareback body but it has certainly an influence on the near wake, which is discussed in the following section. The aerodynamic data are not used in the present study.

\section{Dynamic of the wake barycenter}

\subsection{Wake characterization}

\begin{figure}[h]
\setlength{\unitlength}{1cm}
\includegraphics[width=\linewidth]{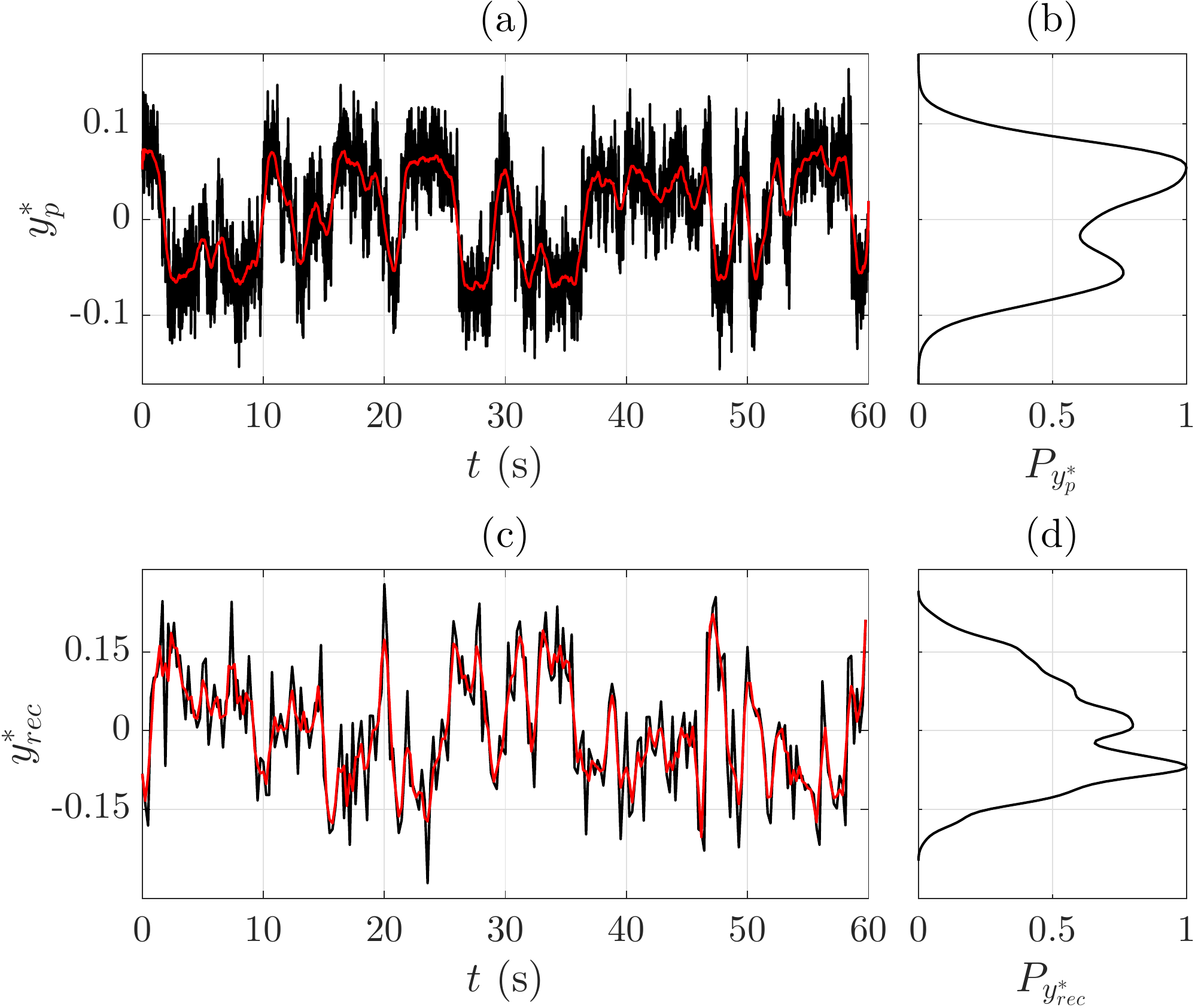}
\caption{Pressure barycenter spanwise position $y_P^*$: (a) evolution in time and (b) normalized PDF $P_{y_p}^*$ for $T_P=1$ min ($f_P=500$ Hz). Recirculation intensity barycenter spanwise position $y_{rec}^*$: (c) evolution in time and (d) normalized PDF $P_{y_{rec}^*}$ for $T_{PIV}=1$ min. Red curves show the data smoothed over one second.}\label{fig:pressure_PIV_yPDF}
\end{figure}

We are interested in the large-scale dynamics of the global wake. This is the reason why we define the instantaneous wall pressure barycenter which can be seen as the footprint of the wake. Denoting the space average of a quantity $f$ over an area $S$ as $\langle f \rangle_S$, we compute the pressure barycenter at $x^*=0$ as:
\begin{equation}
\overrightarrow{O{G_p}(t)} = \left(
\begin{array}{ll}
	y_{p}^*(t) \\
	z_{p}^*(t)
\end{array}
\right) = \left(
\begin{array}{ll}
	\frac{\langle y^*p(t) \rangle_{S_p}}{\langle p(t) \rangle_{S_p}} \\
	\frac{\langle z^*p(t) \rangle_{S_p}}{\langle p(t) \rangle_{S_p}}
\end{array}
\right),
\label{eq:Yp}
\end{equation}
where $p(t)=p(y^*,z^*,t)$ is the local pressure measured at time $t$. Thus the instantaneous barycenter of the depression can be tracked at each time step. In the same way we define the instantaneous recirculation intensity barycenter from the velocity fields in the PIV horizontal plane at $z^*=1$ as:
\begin{equation}
\overrightarrow{OG_{rec}(t)} = \left(
\begin{array}{ll}
	x_{rec}^*(t) \\
	y_{rec}^*(t)
\end{array}
\right) = \left(
\begin{array}{ll}
	\frac{\langle x^*u_{rec}(t) \rangle_{A_{rec}(t)}}{\langle u_{rec}(t) \rangle_{A_{rec}(t)}} \\
	\frac{\langle y^*u_{rec}(t) \rangle_{A_{rec}(t)}}{\langle u_{rec}(t) \rangle_{A_{rec}(t)}}
\end{array}
\right),
\label{eq:YArec}
\end{equation}

where $u_{rec}(t)=u_{rec}(x^*,y^*,t)$ and $A_{rec}(t)$ are respectively the local streamwise component of the recirculation velocity and the recirculation area at time $t$. An example of the evolution in time is given in Fig. \ref{fig:pressure_PIV_yPDF} for $y_p^*$ (a) and $y_{rec}^*$ (c).

The probability density function (PDF) is then computed from the spanwise position of both barycenters. Figure \ref{fig:pressure_PIV_yPDF} shows clearly the bimodality behavior of the turbulent wake whether it is through the pressure barycenter PDF $P_{y_p^*}$ (b)  or the recirculation barycenter PDF $P_{y_{rec}^*}$ (d). Indeed their PDFs have two peaks whose positions are nearly symmetric with respect to the y-axis. The difference in the peaks value is due to the low acquisition time as observed in \cite{Cadot2015}. Longer acquisitions only done for the pressure show well two peaks of the same level. The small amount of data regarding the PIV explains the noisy aspect of $P_{y_{rec}^*}$. The two identified RSB modes are the ($y_p^*<0$, $y_{rec}^*>0$) state and the ($y_p^*>0$, $y_{rec}^*<0$) state. \\
Even if the barycenters seem to switch successively from a mode to the other randomly, some characteristic time scales and characteristic frequencies can be estimated. The mean time spent in each of this mode is $T_{RSB}=1.57 \pm 0.32$ s. The estimated switching frequency is $f_{switch}=0.56 \pm 0.08$ Hz, whereas the switch itself lasts for $T_{switch}=0.30 \pm 0.05$ s. As expected these results have a high dispersion: their respective standard deviations are $\sigma_{T_{RSB}}=2.1 \pm 0.3$ s and $\sigma_{T_{switch}}=0.11 \pm 0.01$ s.

\begin{figure}[h]
\setlength{\unitlength}{1cm}
\includegraphics[width=\linewidth]{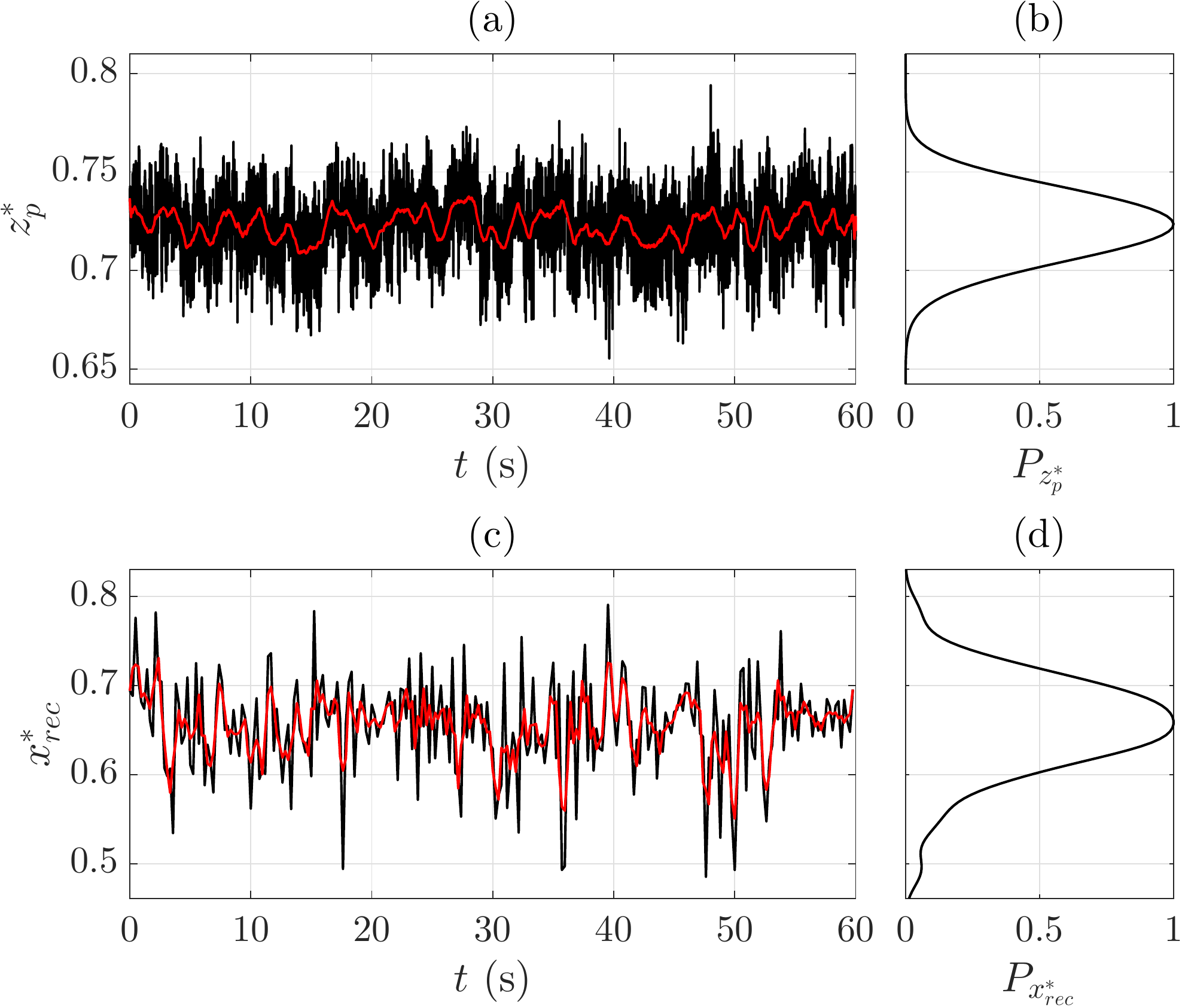}
\caption{Pressure barycenter spanwise position $z_P^*$: (a) evolution in time and (b) normalized PDF $P_{z_p}^*$ for $T_P=1$ min ($f_P=500$ Hz). Recirculation intensity barycenter spanwise position $x_{rec}^*$: (c) evolution in time and (d) normalized PDF $P_{x_{rec}^*}$ for $T_{PIV}=1$ min. Red curves show the data smoothed over one second.}\label{fig:pressure_PIV_xzPDF}
\end{figure}

Regarding the normal to the wall position $z_p^*$ and the streamwise position $x_{rec}^*$ presented in Fig. \ref{fig:pressure_PIV_xzPDF} for the pressure (a-b) and the recirculation intensity respectively (c-d), these positions are stable. Indeed their respective PDF, $P_{z_p^*}$ and $P_{x_{rec}^*}$, show only one mode. 

Thus the same lateral symmetry breaking mechanism is observed in our wake topology as the ones presented without a central leg in \cite{Grandemange2013a,Cadot2015}. The effects of the modified ground clearance on the wake reversal behind a square-back bluff body have been studied by \cite{Barros2015}. It was reported in particular that the presence of a circular cylinder, very similar to the central leg attached to the aerodynamical balance in our experiments, changes only the mode position for $z^*_p$ but does not cancel the lateral bimodality.\\

\begin{figure}[h]
\setlength{\unitlength}{1cm}
\includegraphics[width=\linewidth]{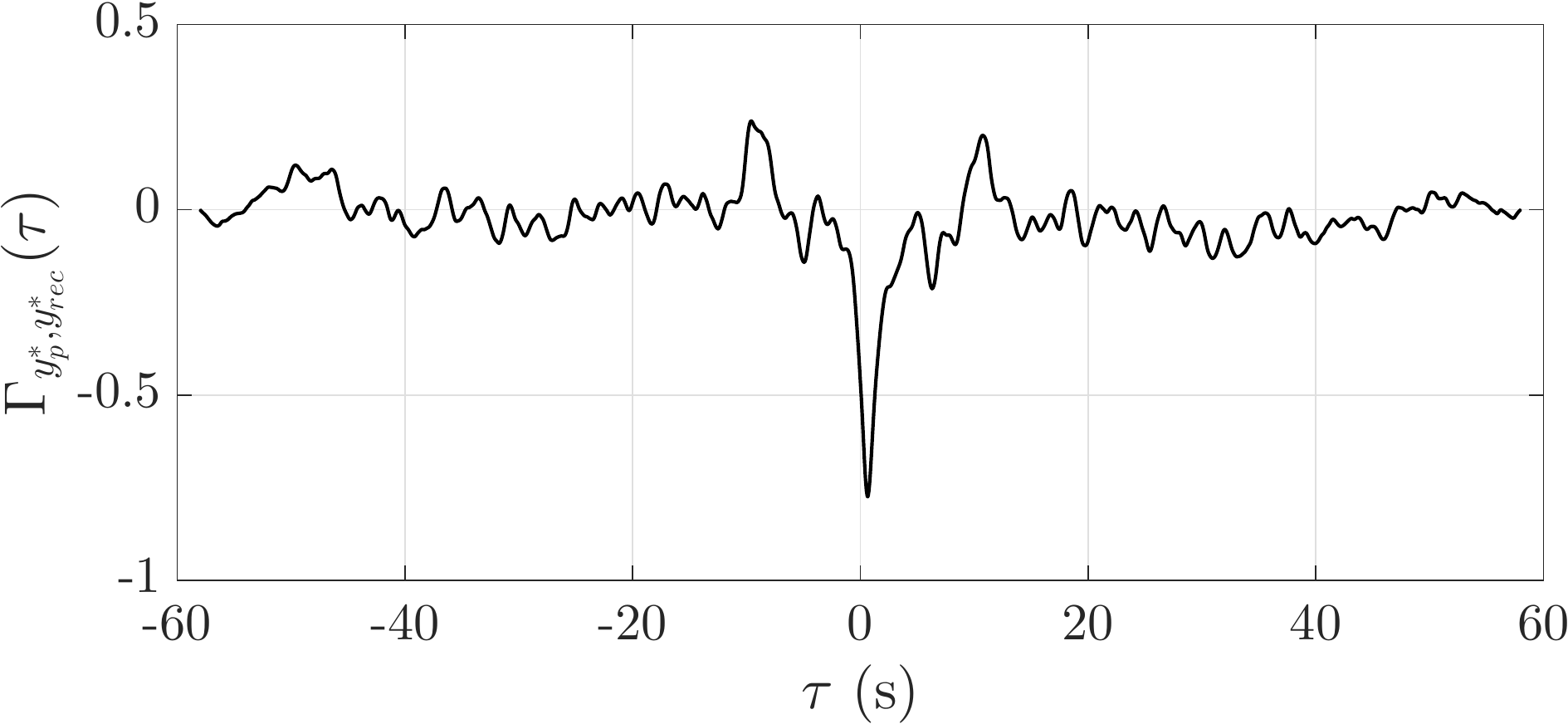}
\caption{Normalized cross-correlation of $y_P^*$ and $y_{rec}^*$.}\label{fig:xcorr_YpYArec}
\end{figure}

The relation between the dynamics of the wake, characterized by its instantaneous recirculation area, and the dynamics wall pressure is not straightforward because of the complex 3D flow. To study their relationship, the normalized cross-correlation of $y_P^*$ and $y_{rec}^*$ has been computed for five one minute runs: ${\Gamma_{y_p^*,y_{rec}^*}(\tau) = y_p^*(\tau) \otimes y_{rec}^*(\tau)}$. The results show a strong correlation and an almost non-existent delay between the two measurements as illustrated by Fig. \ref{fig:xcorr_YpYArec}. 

The value of the peak is ${\Gamma_{y_p^*,y_{rec}^*}(\tau\sim0) = -0.84 \pm 0.05 }$ so both barycenters are in phase opposition. When the depression is located in the negative part of the rear surface ($y_p^*<0$), the recirculation bubble is located in the positive one ($y_{rec}^*>0$) and \textit{vice versa}. This behavior was first observed in \citep{Grandemange2012b}. So the wake states can be characterized either by the pressure barycenter or by the recirculation intensity one. In the following we only present results regarding $y_p^*$ for several independent experiments of various lengths $T_P=\lbrace 2; 5 \rbrace$ min acquisition runs ($f_P=250$ Hz and $f_P=100$ Hz respectively).

\subsection{Coherent structures}

\begin{figure}[h]
\setlength{\unitlength}{1cm}
\includegraphics[width=\linewidth]{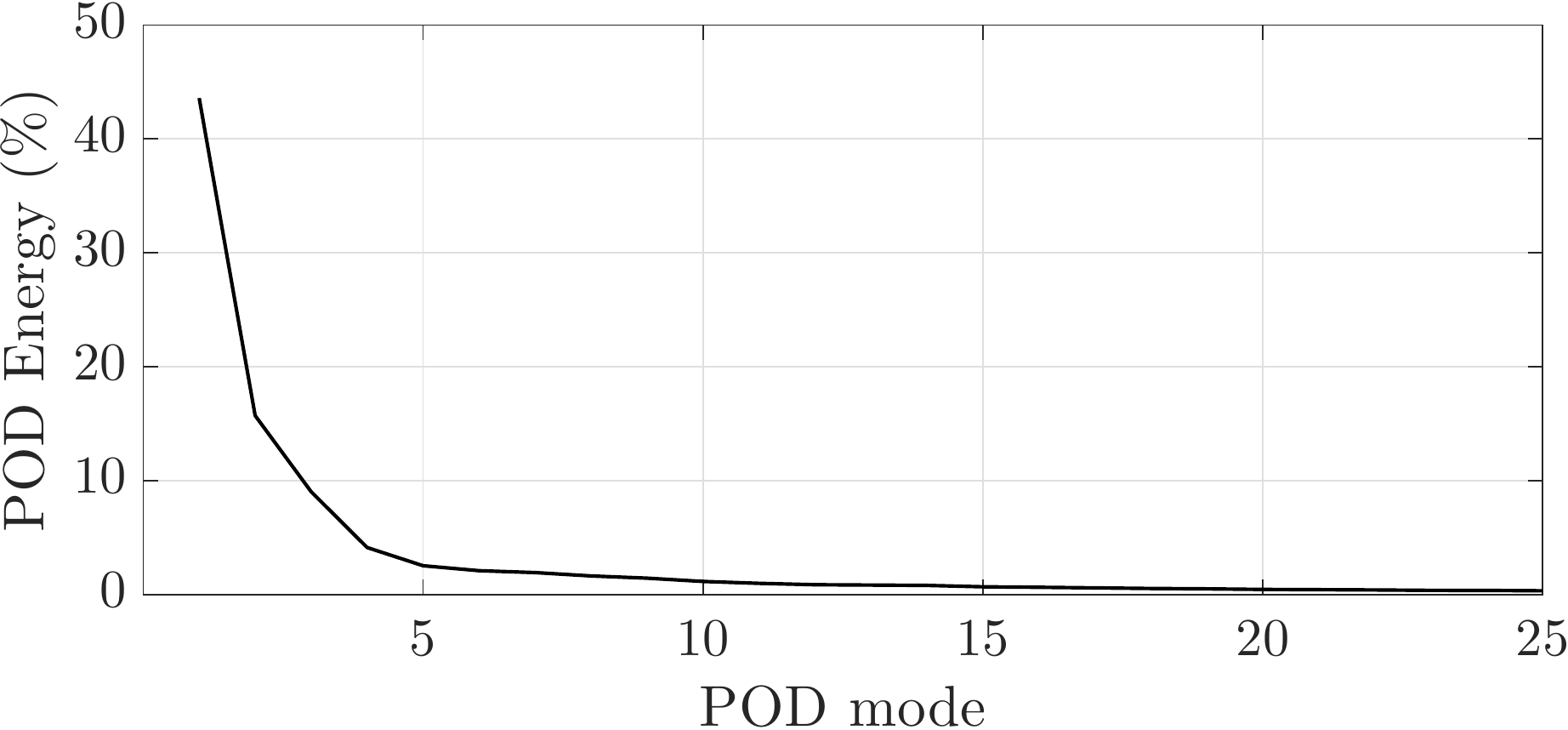}
\caption{Energy distribution of the first 25 POD modes of the rear pressure.}\label{fig:PODenergy}
\end{figure}

As we are interested in the dynamics of the large-scale structures, we analyze the spatio-temporal organization of the wall pressure spatial distributions using the proper orthogonal decomposition (POD). It is an efficient approach to detect coherent structures in turbulent flows \cite{Lumley1967,Sirovich1987}. Thus we apply the POD on the rear  pressure coefficient fluctuations $\tilde{C_p}$:
\begin{equation}
\tilde{C_p}(t) \sim \sum_{i=1}^{k} a_i(t)\Phi_i,
\end{equation}
where $k$ is the number of POD modes $\Phi_i$ carrying most of the coherent structures energy and $a_i$ are the corresponding temporal coefficients. The energy distribution of the first 25 POD modes is given on Fig. \ref{fig:PODenergy} showing that the first five modes contain $75\%$ of the total energy.

The power spectral density (PSD) is computed to analyze the dynamics of the first POD modes. For practical comparisons we express the frequency $f$ as the Strouhal number based on the width bluff body: $St_W=fW/U_\infty$.

\begin{figure}[h]
\setlength{\unitlength}{1cm}
\includegraphics[width=\linewidth]{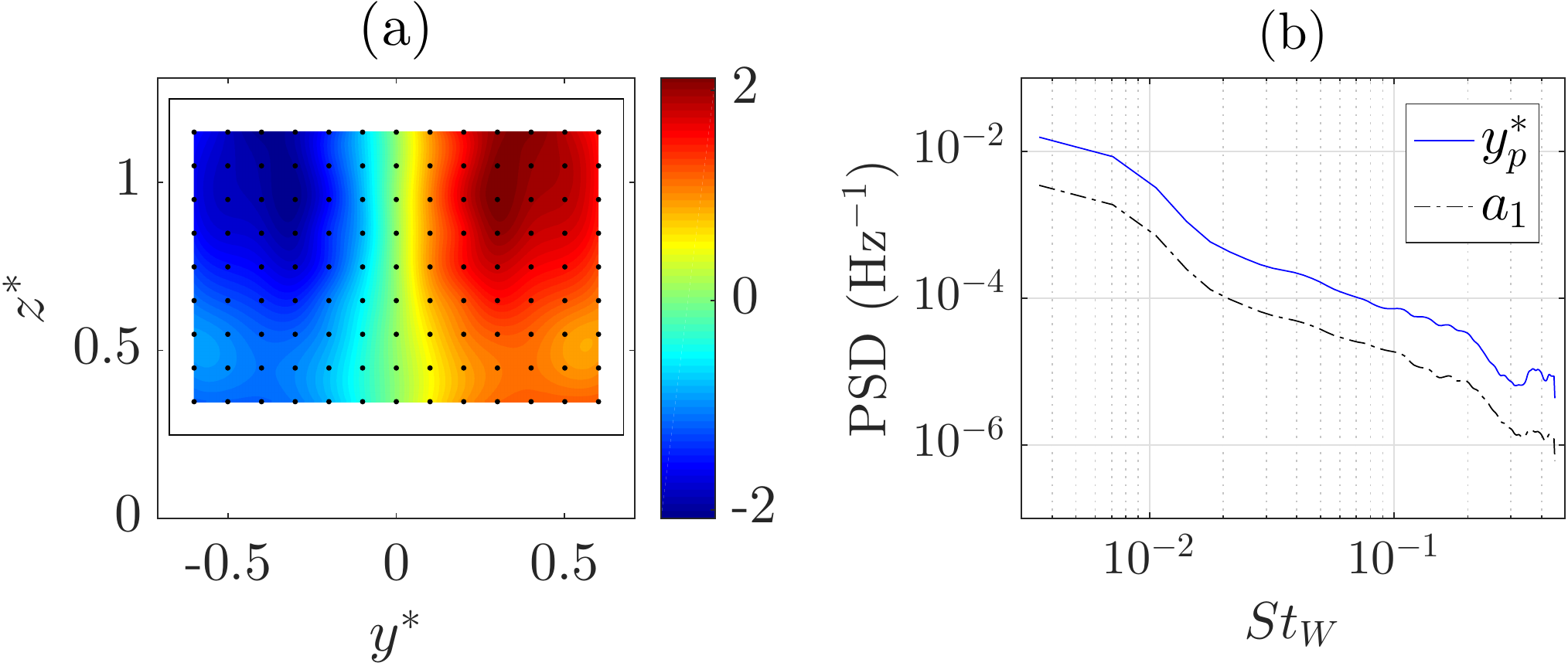}
\caption{(a) First POD mode $\Phi_1$ ($43\%$) and (b) PSD of its corresponding coefficient $a_1$ (dashed red line) together with the PSD of $y_p^*$ (solid blue line).}
\label{fig:PODmode1}
\end{figure}

The most remarkable results are that the first mode $\Phi_1$ represents $43\%$ of the total energy and that its spatial organization exhibits the global symmetry breaking presented on Fig. \ref{fig:PODmode1} (a). Moreover its spectral signature is identical to the one of the pressure barycenter, displayed on Fig. \ref{fig:PODmode1} (b): the low frequencies, $St_W<0.02$, contain most of the power spectrum. Thus the pressure barycenter is a direct measure of the most energetic large-scale coherent structure governed by a long time scale.

\subsection{Quasi strange attractor dynamics}

\begin{figure}[h]
\setlength{\unitlength}{1cm}
\includegraphics[width=\linewidth]{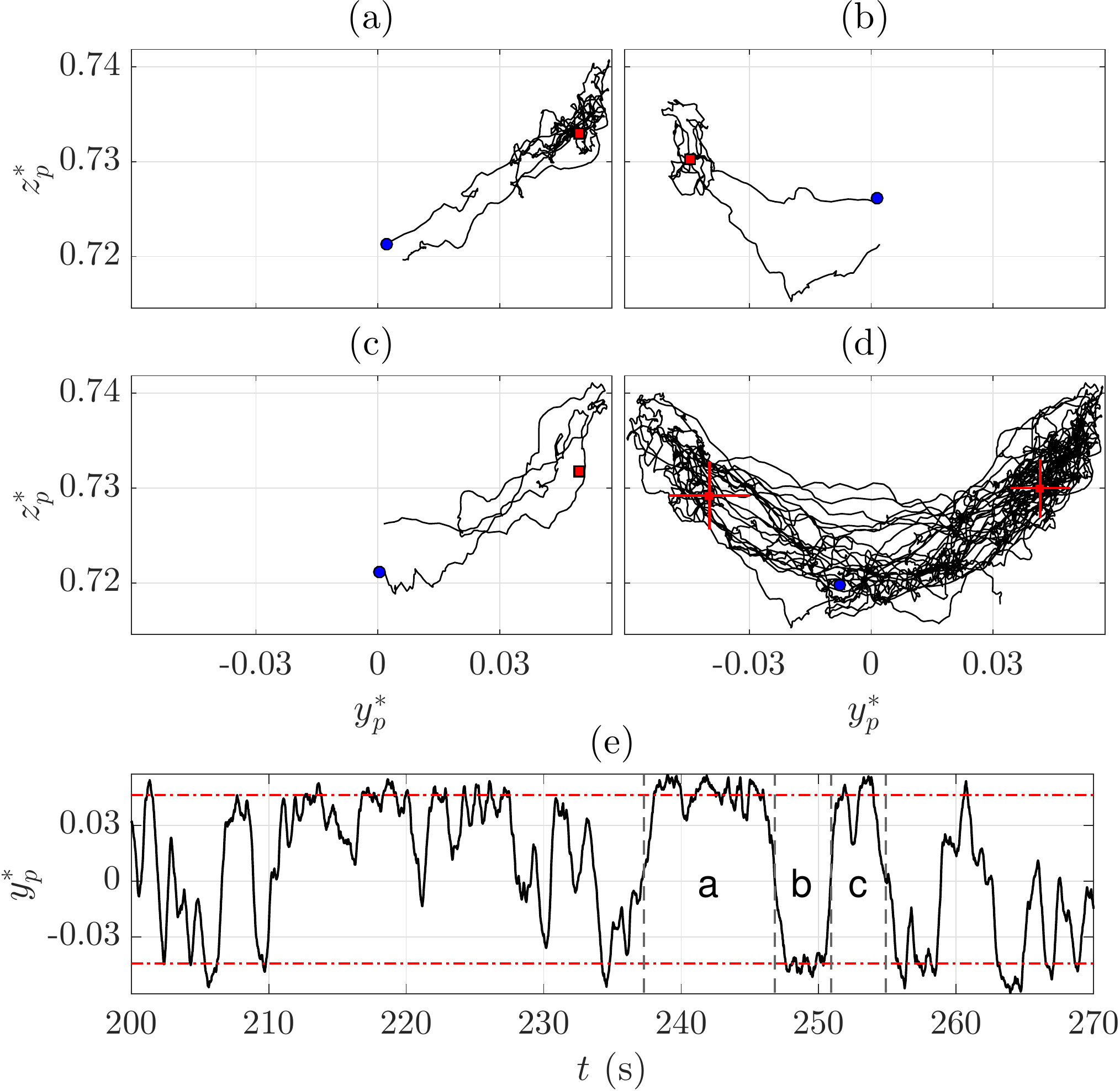}
\caption{Evolution in space of the pressure barycenter $G_P$ (black line for the trajectory and blue circle for the last position) with the identified centers (red squares) for three successives small intervals: (a) $t \in [237;247]$, (b) $t \in [247;251]$, (c) $t \in [251;255]$ and (d) a larger interval  $t \in [237;266]$ (red cross for the standard deviation). (e) $y_p^*(t)$ for $t \in [200;270]$ with the spanwise position of the two centers (dash-dotted red lines). All intervals are in seconds and data are smoothed for clarity. Movie available online.}\label{fig:ZpYp_evolution}
\end{figure}

The positions of the pressure barycenter $G_P$ for three successive stays in the RSB modes are displayed on Fig. \ref{fig:ZpYp_evolution} (a-b-c). We also plot the most probable position (red square) near which $G_p$ evolves during each stay.

Figure \ref{fig:ZpYp_evolution} (d-e) show longer tracking with the most probable positions of each RSB mode, disclosing two $y$-axis symmetric points which acts like the centers of a strange attractor.\\
The two dimensional PDF of the pressure barycenter is computed to analyze its most frequent positions on the rear. We also calculate the positions of the identified centers, denoted $C_1 (y_{C_1}^*,z_{C_1}^*)$ and $C_2 (y_{C_2}^*,z_{C_2}^*)$, as local modes of the PDF $P_{y_p^*}$, such as: 
\begin{equation}
\overrightarrow{OC_1} = \left(
\begin{array}{ll}
	 {y_p^*}_{ \vert \underset{y_p^*<0}{\max} P_{y_p^*} }\\
	 {z_p^*}_{ \vert \underset{y_p^*<0}{\max} P_{y_p^*} }
\end{array}
\right)~\textrm{and}~
\overrightarrow{OC_2} = \left(
\begin{array}{ll}
	 {y_p^*}_{ \vert \underset{y_p^*>0}{\max} P_{y_p^*} }\\
	 {z_p^*}_{ \vert \underset{y_p^*>0}{\max} P_{y_p^*} }
\end{array}
\right).
\label{eq:attractorCenter}
\end{equation}

\begin{figure}[h]
\setlength{\unitlength}{1cm}
\includegraphics[width=\linewidth]{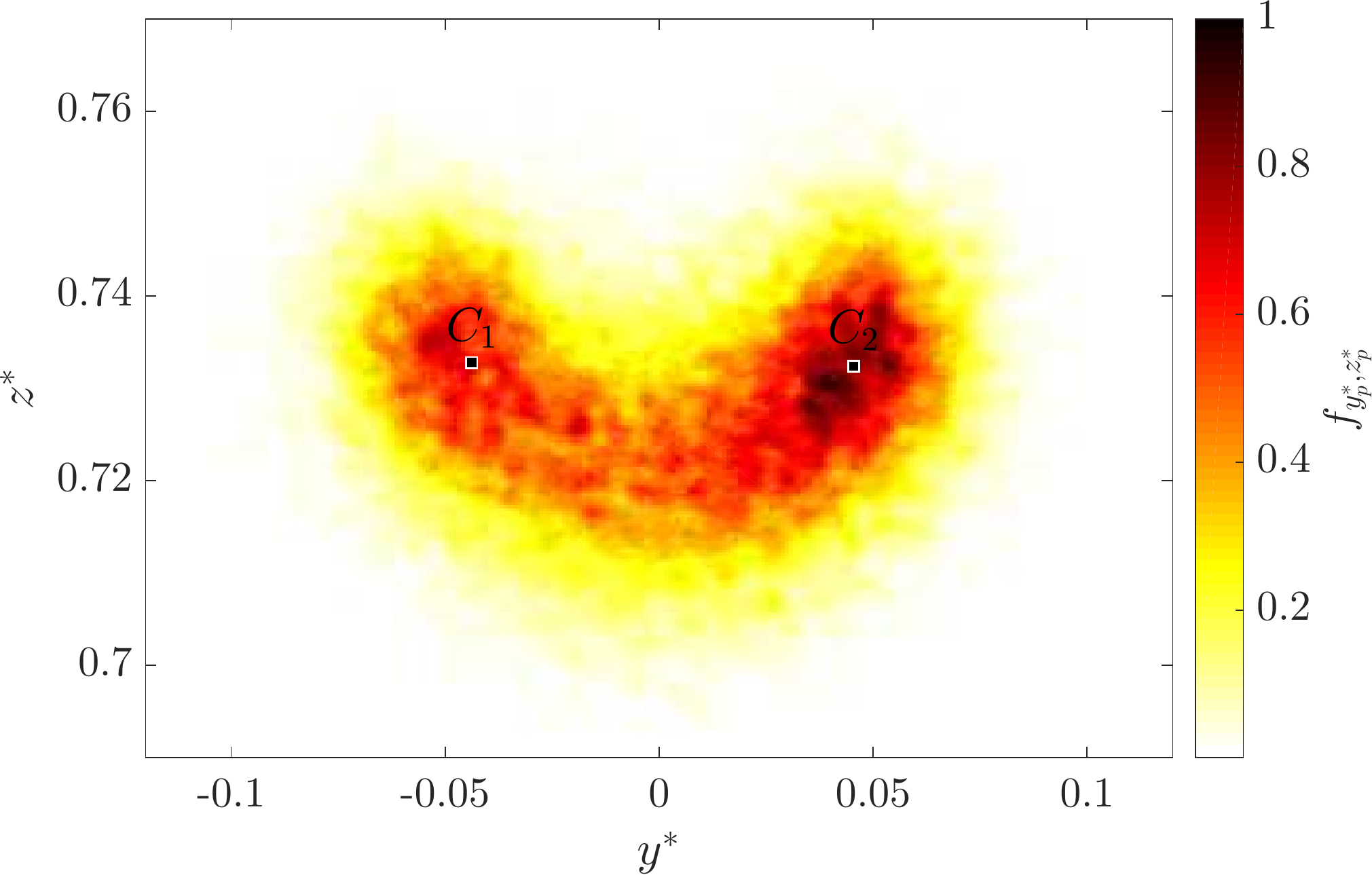}
\caption{Normalized two dimensional PDF of the pressure barycenter $G_P$ with the two identified local maxima of the PDF (black squares).}\label{fig:pressure_pdf2D}
\end{figure}

The results, shown in Fig. \ref{fig:pressure_pdf2D}, reveal two areas highly preferred by the barycenter, corresponding to red the two RSB states. For each area a quasi-attractive center can be identified, such as $y_{C_1}^* \sim - y_{C_2}^*$ and $z_{C_1}^* \sim z_{C_1}^*$. During the switch between these two positions, the pressure barycenter follows preferentially a trajectory along a well-defined path. In the following section we characterize the signal $y_p^*$.

\section{Characterization of the attractor}

\subsection{Structure function}
\begin{figure}[h]
\setlength{\unitlength}{1cm}
\includegraphics[width=\linewidth]{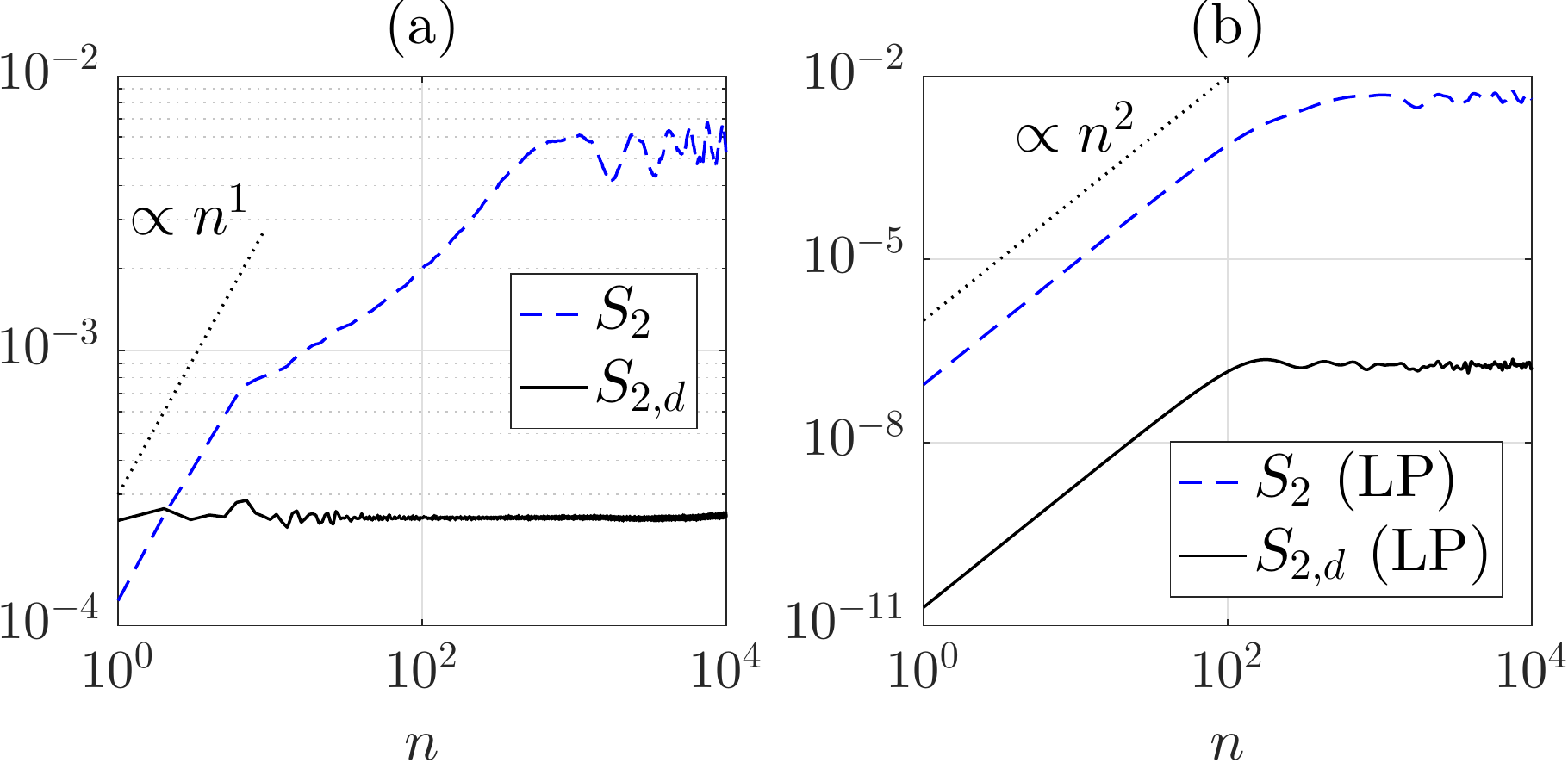}
\caption{(a) Second order structure functions of $y_p^*$ (dashed blue line) and of its first derivative (solid black line). (b) Second order structure functions of the LP filtered $y_p^*$ (dashed blue line) and of its first derivative (solid black line). $f_p=250$ Hz.}
\label{fig:structFunction}
\end{figure}

Analyzing the signal to know whether the dynamics of the wake are chaotic or stochastic is of prime interest. An effective approach is to study its self-affinity by computing its first order ($k=1$) \cite{Osborne1989} or its second order ($k=2$) \cite{Provenzale1992} structure function $S_k$, defined as:

\begin{equation}
S_k(n)=\langle \vert y_p^*(i+n)-y_p^*(i) \vert^k \rangle_i, 
\end{equation}

where $n$ is the lag and $\langle . \rangle_i$ stands for the average over $N-n$ points. According to \cite{Mandelbrot1982}, if $y_p^*$ is fractal, $S_k$ follows a scaling law for small $n$:

\begin{equation}
S_k(n) \propto n^{k h}, 
\end{equation}

where $h$ is the so-called scaling exponent. If the signal is chaotic then $h=1$. If it is stochastic then its power spectrum follows a power-law $PSD(St_W) \propto St_W^{-\alpha}$ and $\alpha = 2h+1$.\\
However, $S_k$ alone is not enough to conclude and the structure function of the first derivative of the signal $y_p^*$, denoted as $S_{k,d}$, needs to be computed \cite{Provenzale1992}. Thus, if it is stochastic then $S_{k,d}$ is almost constant and if the signal is chaotic then $S_{k,d}(n)$ follows a scaling law for small $n$. Figure \ref{fig:structFunction} displays $S_2$ and $S_{2,d}$ for the raw signal $y_p^*$ (a) and for the signal on which we apply a low-pass (LP) filter (b) to only select low frequencies modes which have the highest magnitudes: $St_W<10^{-3}$ (see Fig. \ref{fig:PODmode1} (b)).

The raw signal behaves clearly like a fractal noise with $h=1/2$ implying $\alpha=2$, which is in agreement with \citep{Grandemange2013a,Brackston2016}. Conversely the LP filtered data appear to be chaotic since $h=1$ and $S_{2,d}$ is not constant. It is noteworthy that the same observations hold for $k=1$.

\subsection{Embedding dimension}

As we only observe a part of the non-linear wake dynamics (rear body pressure and 2D-2C velocity fields of the near-wake), we cannot experimentally access to its complete phase space. But \cite{Eckmann1985} and \cite{Sauer1991} provide the so-called embedding methods to reconstruct a pseudo phase space from time series. Thus, according to the pioneering Takens' time-delay embedding method \cite{Takens1981}, we build the $M$ state vectors $\lbrace Y_i \rbrace_{i=1...M}$ :
\begin{equation}
Y_i(m)= \left[~y^*_p(i)~y^*_p(i+J)~...~y^*_p(i+(m-1)J)~\right],
\end{equation}

where $m$ is the embedding dimension and $J$ the reconstruction delay, implying $M=N-(m-1)J$. $J$ is computed trough the mutual information process \cite{Fraser1986} using improved kernel density estimation algorithm \cite{Silverman1986,Thomas2014} to avoid \textit{redundance} ($J$ too small) and \textit{irrelevance} ($J$ too large) in phase space reconstruction \cite{Casdagli1991}. We then determine the minimum value of $m$ from $y_p^*$, following Cao's method \cite{Cao1997}, which is based on the false nearest neighbor algorithm of \cite{Kennel1992}. It should be noted that the values obtained for $J$ are also checked through improved Celluci's algorithm \cite{Cellucci2005,Jiang2010}.

Cao's method is summarized here since it enables also to distinguish deterministic and stochastic time series. The idea is to evaluate how the mean distance between close state vectors $E(m)$ evolves with respect to $m$. Over the real embedding dimension $E(m)$ does not change anymore. We look for the closest neighbor $Y_{\mathcal{N}(i,m)}(m)$ of each state vector $Y_i(m)$: 
\begin{equation}
\underset{\mathcal{N}(i,m) \neq i}{\min} \Vert Y_i(m)-Y_{\mathcal{N}(i,m)}(m) \Vert,~i \in [\![ 1;M ]\!],
\label{eq:neighbor}
\end{equation}
where $\Vert . \Vert$ stands for the Euclidean distance. The ratio of the distances in $m$ and $m+1$ dimensions for $Y_i$ is: 
\begin{equation}
a_{i,m}=\frac{\Vert Y_i(m+1)-Y_{\mathcal{N}(i,m)}(m+1) \Vert}{\Vert Y_i(m)-Y_{\mathcal{N}(i,m)}(m) \Vert}.
\end{equation}
Finally $E(m)$ is computed:
\begin{equation}
E(m)=\frac{1}{M}\sum_i a_{i,m}.
\end{equation}
For convenience its evolution is evaluated through $E_1(m)$:
\begin{equation}
E_1(m)=\frac{E(m+1)}{E(m)},
\label{eq:E1}
\end{equation}
and if $\exists~m~\backslash~\forall k \geq m,~E_1(k+1)=E_1(k)$ then $m$ is the minimum embedding dimension. In parallel the mean difference between the raw data $E^*(m)$ is evaluated with respect to $m$, relating to $\mathcal{N}(m)$ obtained in Eq. (\ref{eq:neighbor}):
\begin{equation}
E^*(m)=\frac{1}{M}\sum_i \vert y_p^*(i+mJ)-y_p^*(\mathcal{N}(i,m)+mJ) \vert.
\end{equation}
In the same manner as the Eq. (\ref{eq:E1}), the evolution is analyzed through $E_2(m)$:
\begin{equation}
E_2(m)=\frac{E^*(m+1)}{E^*(m)},
\end{equation}
and if $\exists~k~\backslash~E_2(k) \neq 1$ then the signal is deterministic, otherwise it is stochastic.

\begin{figure}[h]
\setlength{\unitlength}{1cm}
\includegraphics[width=\linewidth]{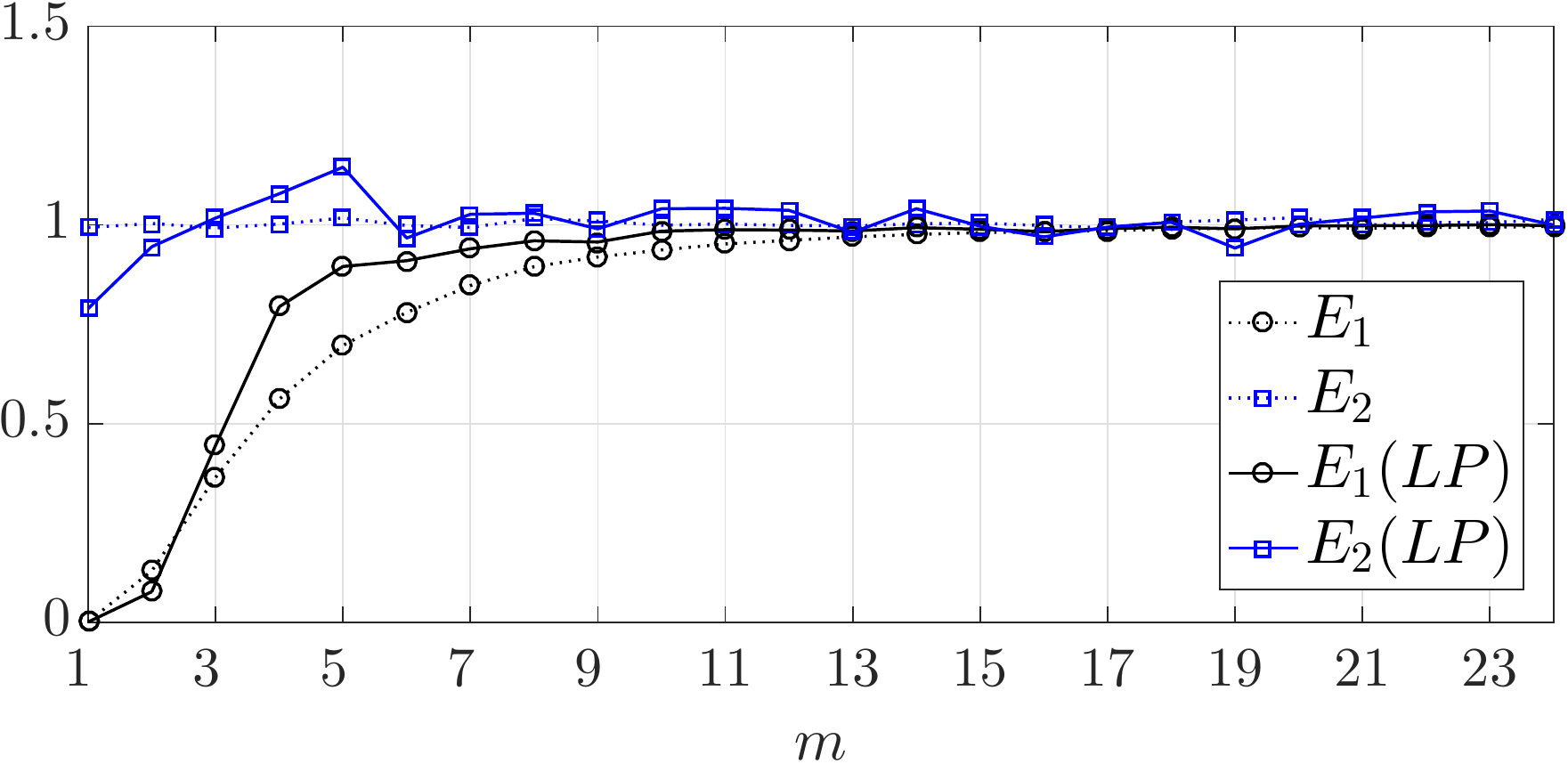}
\caption{$E_1(m)$ (black circle) and $E_2(m)$ (blue square) computed from raw (dashed line) and LP filtered (solid line) signal $y_p^*$. $f_p=100$ Hz. The large-scale dynamics associated with the LP filtered signal show a chaotic signature.}
\label{fig:E1_E2}
\end{figure}

As shown by Fig. \ref{fig:E1_E2}, the raw signal appears stochastic but the LP filtered one is indeed deterministic, confirming the results obtained with the structure function. The minimum embedding dimension seems to be $m=19$. It should be noted that a new delay $J$ is computed for the filtered data. \\
Thus the complete large-scale bimodal dynamics of such a fully turbulent wake are well composed of a stochastic part and a deterministic part as recently proposed by \citep{Brackston2016}. In the following sections we characterize the chaotic behavior of these dynamics through the analysis of the LP filtered signal directly denoted as $y_p^*(t)$.\\

\subsection{Largest Lyapunov exponent}

There are many ways to characterize and quantify chaos. Among the most popular quantities, one can cite the correlation dimension \cite{Grassberger1983} which gives an estimate of the system complexity and characteristic exponents which give an estimate of the level of chaos in the dynamical system. In this section we focus on the Lyapunov exponent. The spectrum of Lyapunov exponents is well known for detecting and quantifying chaotic systems from the experimental time series \cite{Wolf1985}. Indeed chaos exists if a system is sensitive to its initial conditions. Thus the principle consists in following the evolution of the distance $d$ between two initial neighboring state vectors in the phase space. For a chaotic attractor, the distance $d$ exponentially increases in time at an average rate equal to the largest Lyapunov exponent (LLE) $\lambda_1$ \cite{Eckmann2004}:
\begin{equation}
d(t)=d(0)\exp^{\lambda_1t}.
\label{eq:chaosDistance}
\end{equation} 

We apply Rosenstein's algorithm \cite{Rosenstein1992,Sato1987}, rather than Wolf's algorithm \cite{Wolf1985}, to our LP filtered $N$ points time series $\lbrace y^*_p(i) \rbrace_{i=1...N}$, due to its efficiency on small data sets. We use the $M$ state vectors $\lbrace Y_i \rbrace_{i=1...M}$ and their closest neighbor $\lbrace Y_{\mathcal{N}(i)} \rbrace_{i=1...M}$ as previously. Then we compute the distances evolution:
\begin{equation}
d_i(j)= \Vert Y_{i+j}-Y_{\mathcal{N}(i)+j} \Vert,
\end{equation}

where $j=[\![ f_pt ]\!]$ verifies $i+j \leq M$ and $\mathcal{N}(i)+j \leq M$. The distance $d(t)$ is approximated by averaging over $i$ the distances $d_i(j)$:

\begin{equation}
d(j)= \langle d_i(j)\rangle_i.
\label{eq:finalDistance}
\end{equation}

\begin{figure}[h]
\setlength{\unitlength}{1cm}
\includegraphics[width=\linewidth]{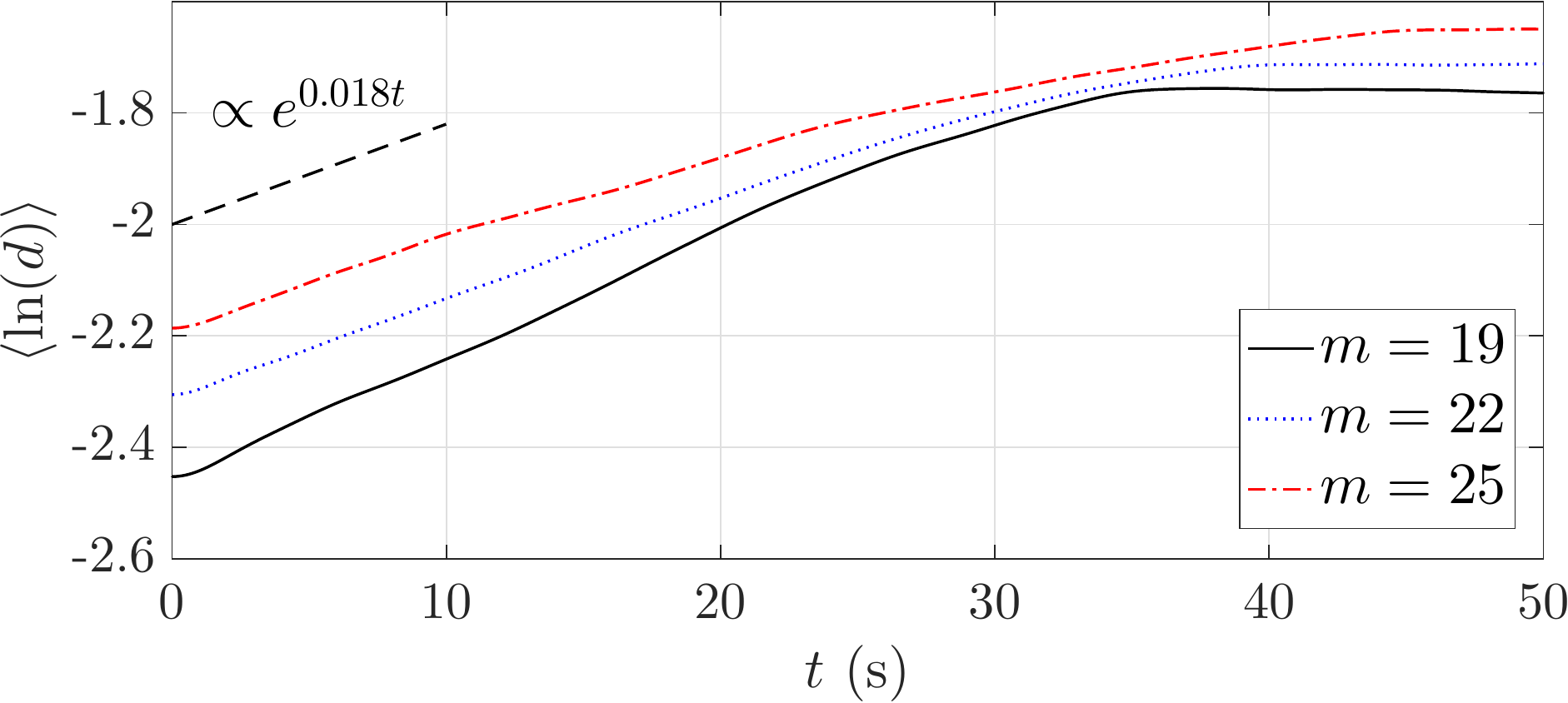}
\caption{$\langle ln(d(t)) \rangle_{i=1..M}$ and exponential approximation $e^{0.018 t}$ (black dashed line). $f_p=100$ Hz.}
\label{fig:LLEslope}
\end{figure}

By taking the logarithm of Eq. (\ref{eq:finalDistance}) with respect to $j$ the slope, extracted through a least-square fit, gives directly the LLE. The distance $d(t)$ computed for our LP filtered data is given in Fig. \ref{fig:LLEslope} and it appears to follow the Eq. (\ref{eq:chaosDistance}) for small $t$, typically $t \in \lbrack 0;1 \rbrack$ (in second). The LLE is thus computed in this range.

The estimated positive LLE is $\lambda_1=0.018 \pm 0.003$ s$^{-1}$. The relative error is due to the main difficulty to define the right linear region of the curve to fit. The low frequencies dynamics associated to large-scale structures can thus be considered as a weak chaotic strange attractor.

\subsection{Correlation dimension}

The fractal dimension (or Hausdorff dimension) $D$ of a strange attractor can be rigorously approximated by its correlation dimension $D_2$ which is directly computed from experimental time series according to the works of \cite{Grassberger1983}. The previous $M$ state vectors $\lbrace Y_i \rbrace_{i=1...M}$ are also used to compute the correlation integral function $\mathcal{C}(\tau)$, defined as:

\begin{equation}
\mathcal{C}(\tau) = \frac{2}{M(M-1)} \sum_{i=1}^{M} \sum_{j=i+1}^{M} \Theta (\tau-\Vert Y_i - Y_j \Vert),
\label{eq:correlInt}
\end{equation}

where $\Theta$ is the Heaviside function. The correlation dimension $D_2$ can be derived from the correlation integral function $\mathcal{C}(\tau)$ which scales as a power-law for small $\tau$:

\begin{equation}
\mathcal{C}(\tau) \propto \tau^{D_2},
\end{equation}

\begin{figure}[h]
\setlength{\unitlength}{1cm}
\includegraphics[width=\linewidth]{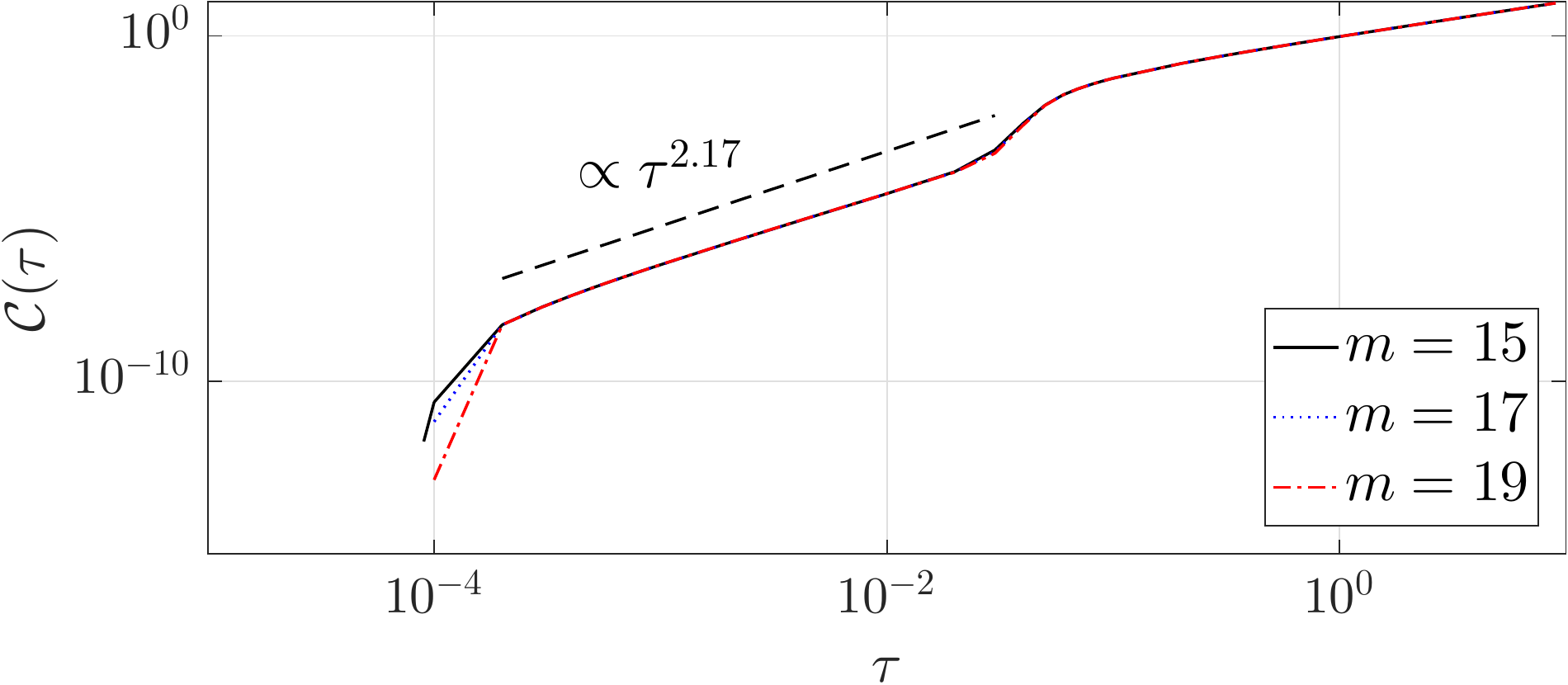}
\caption{Correlation integral function $\mathcal{C}(\tau)$ of LP filtered $y_p^*$ and power law approximation $\propto \tau^{2.17}$ (black dashed line) in logarithmic scale.}
\label{fig:grassberger}
\end{figure}

The correlation integral function of our LP filtered signal $y_p^*$ is computed using the Grassberger-Procaccia method improved by \cite{Ning2008} who normalize the Euclidean distance in Eq. (\ref{eq:correlInt}) by the embedding dimension $m$. Figure \ref{fig:grassberger} shows the resulting correlation integral function in logarithmic scale.

Regarding the whole experimental signals we obtain $D_2=2.17 \pm 0.01$, which verifies $D_2 \leq 2 \log N$ \cite{Ruelle1990}. The correlation dimension is computed over two decades. Furthermore we verify again the low-dimensional aspect of the dynamics using the phase randomization test provided by \cite{Provenzale1992}: for a stochastic signal the correlation dimension does not change when its Fourier phases are randomized. 

\begin{figure}[h]
\setlength{\unitlength}{1cm}
\includegraphics[width=\linewidth]{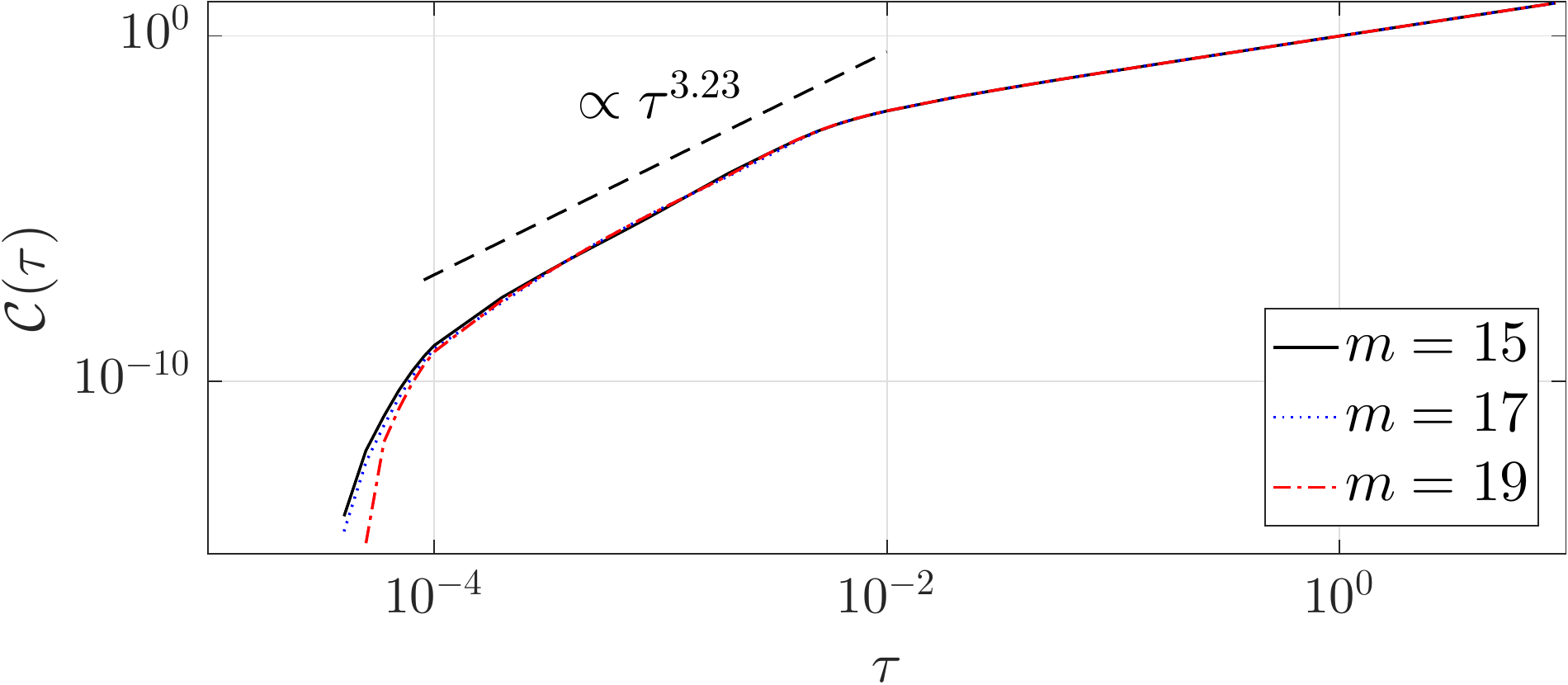}
\caption{Correlation integral function $\mathcal{C}(\tau)$ of LP filtered $y_p^*$ after randomizing its Fourier phases and power law approximation $\propto \tau^{3.23}$ (black dashed line) in logarithmic scale.}
\label{fig:grassbergerRandom}
\end{figure}

Displayed in Fig. \ref{fig:grassbergerRandom}, the computed correlation dimension after phase randomizing of our signal is clearly different: $D_{2,rand}=3.23 \pm 0.01 $, which is indicative of chaotic dynamics.

\subsection{Telegraph-like signal}

Another interesting way to characterize chaotic oscillator is to analyze the properties of its autocorrelation function (ACF). The ACF of a given function $f(t)$ is defined as ${\Gamma_{f}(\tau) = \langle f(t)f(t + \tau) \rangle_T }$, where $\langle . \rangle_T$ stands for the time-averaging. The ACF is normalized by $\Gamma_{f}(0)$. The ACF of a random process has different properties depending on the nature of the system. Among the most popular models used to describe the behavior of many applied random systems, one can cite the noisy harmonic oscillations and the telegraph signal \cite{Anishchenko2002}. The model of telegraph signal is particularly well-suited to describe the statistics of random switching of a bimodal system in the presence of noise, which is a close description of the bimodal wake. One can distinguish two main kinds of telegraph signals, namely, the random and quasi-random telegraph signals. The random telegraph signal
is characterized by a Poisson distribution of switching moments while quasi-random telegraph signal corresponds to random switching between two equi-probable states (probability of switching events equal to $\frac{1}{2}$). For instance, the latter is very well suited to the Lorenz attractor \cite{Anishchenko2002}.

To characterize our time-series as telegraph signals, the first step is then to define two states as the symmetric spanwise positions of the pressure centers as $y^*=\pm y^*_{C_2}$. From our time-series, it is then possible to construct a telegraph-like signal $Y_p^*$ such as:

\begin{equation}
Y_p^*(t) = 
 \begin{cases}
  - y^*_{C_2} & \text{if } y_p^*(t) \leq 0 \\
  y^*_{C_2} & \text{if } y_p^*(t) > 0 
 \end{cases}.
\label{eq:telegraph}
\end{equation}

\begin{figure}[h]
\setlength{\unitlength}{1cm}
\includegraphics[width=\linewidth]{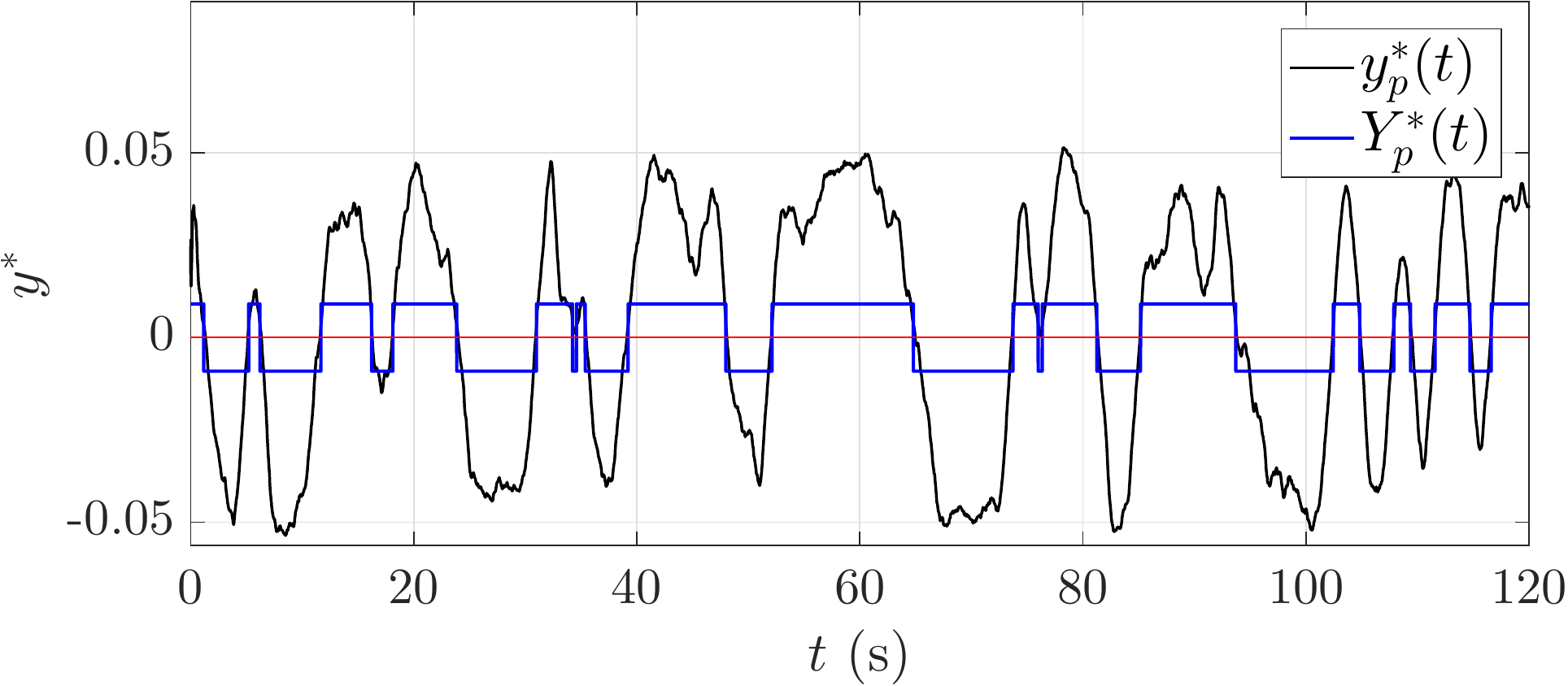}
\caption{Part of the telegraph signal $Y_p^*(t)$ (blue) obtained for the pressure barycenter spanwise position  $y_p^*(t)$ (black). For clarity the telegraph signal is plotted with a 0.25 factor. Here, $y^*_{C_2}=3.64 \times 10^{-2} \pm 1 \times 10^{-4}$.}\label{fig:Telegraph}
\end{figure}

Figure \ref{fig:Telegraph} shows a part of the LP filtered signal $y^*(t)$ together with the resulting telegraph signal $Y_p^*(t)$.

\begin{figure}[h]
\setlength{\unitlength}{1cm}
\includegraphics[width=\linewidth]{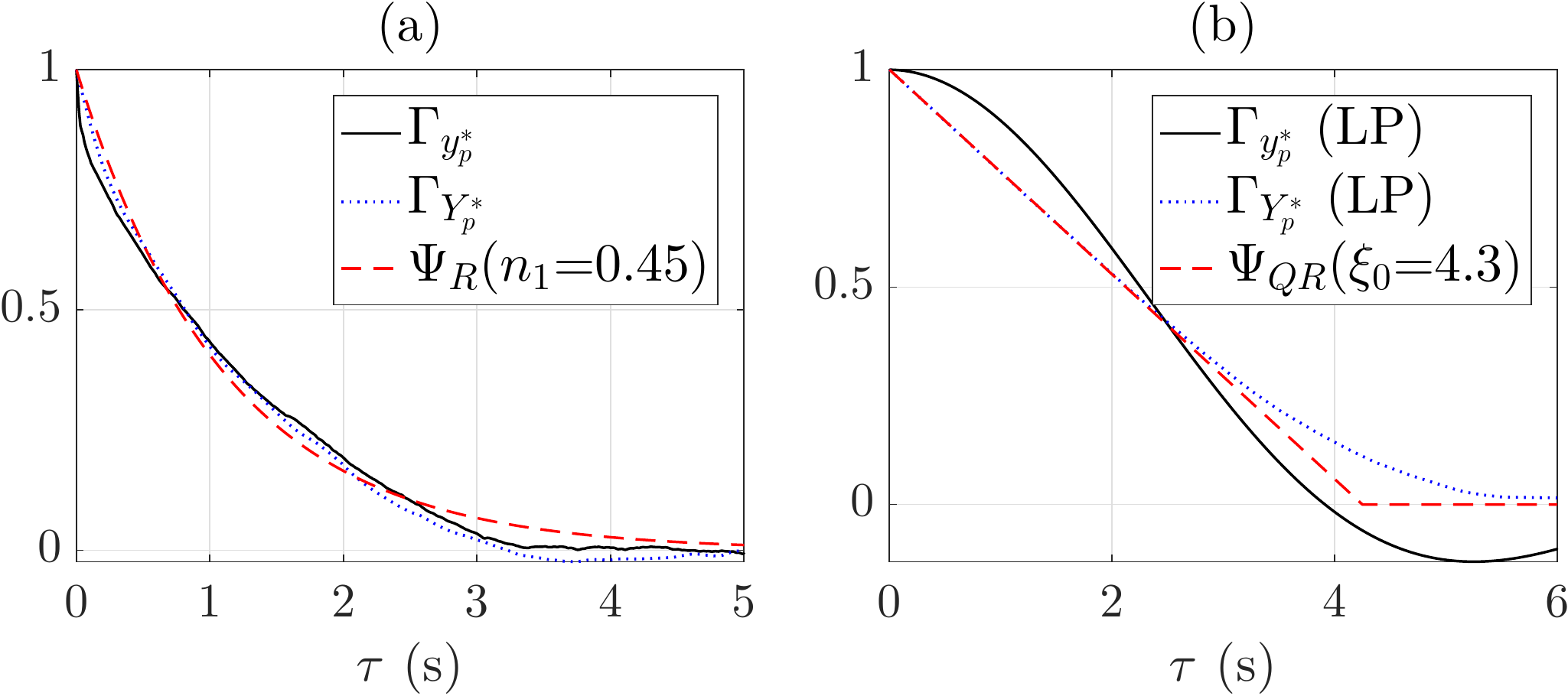}
\caption{Normalized autocorrelation functions of the pressure barycenter spanwise position $\Gamma_{y_p^*}$ (black solid line) and of the resulting telegraph-like signal $\Gamma_{Y_p^*}$ (blue dotted line) for (a) the raw signal and (b) the LP filtered one, together with their respective ACF approximations: $\Psi_R(n_1=0.45)$ and $\Psi_{QR}(\xi_0=4.3)$ (red dashed line).}\label{fig:autoCorr_YpTelegraph}
\end{figure}

The normalized ACF $\Gamma_{y_p^*}(\tau)$ and $\Gamma_{Y_p^*}(\tau)$ are computed for the raw signal and the LP filtered one. According to \cite{Anishchenko2002}, the ACF approximation of a random telegraph signal, $\Psi_R(\tau)$, is given by by the following function : 

\begin{equation}
\Psi_R(\tau)=e^{-2 n_1 \vert \tau \vert},
\label{eq:linearApproxTelegraphR}
\end{equation}

where $n_1$ corresponds to the mean switching frequency, while the ACF of a random telegraph signal can be linearly approximated on short times by the following function $\Psi_{QR}(\tau)$: 

\begin{equation}
\Psi_{QR}(\tau) = 
 \begin{cases} 
  1-\frac{\lvert \tau \rvert}{\xi_0} & \text{if } \lvert \tau \rvert < \xi_0 \\
  0       							  & \text{if } \lvert \tau \rvert \geq \xi_0
 \end{cases},
\label{eq:linearApproxTelegraphQR}
\end{equation}

where $\xi_0$ corresponds to the minimal residence time in one state, denoted $T_{min}$. $\Gamma_{y_p^*}(\tau)$, $\Gamma_{Y_p^*}(\tau)$ and the ACF approximations are plotted in Fig. \ref{fig:autoCorr_YpTelegraph}.

Figure \ref{fig:autoCorr_YpTelegraph} (a) shows that $\Psi_R(n_1=0.45)$ is a good approximation of the ACF of the telegraph signal obtained from the raw data. The computed mean switching frequency $n_1=0.45$ Hz is close to the measured one $f_{switch}=0.56$ Hz. One can see on Fig. \ref{fig:autoCorr_YpTelegraph} (b) that, the ACF of the telegraph-like signal extracted from the LP filtered time-series is well approximated by the linear function $\Psi_{QR}(\xi_0=4.3)$. From the linear approximation we find $\xi_0=4.3$ s, which is similar to our filtered experimental results $T_{min}=3.6$ s. The same ACF approximation is obtained for the Lorenz attractor described by Eq. \ref{eq:LorenzDef} \citep{Anishchenko2002}.

\begin{figure}[h]
\setlength{\unitlength}{1cm}
\includegraphics[width=\linewidth]{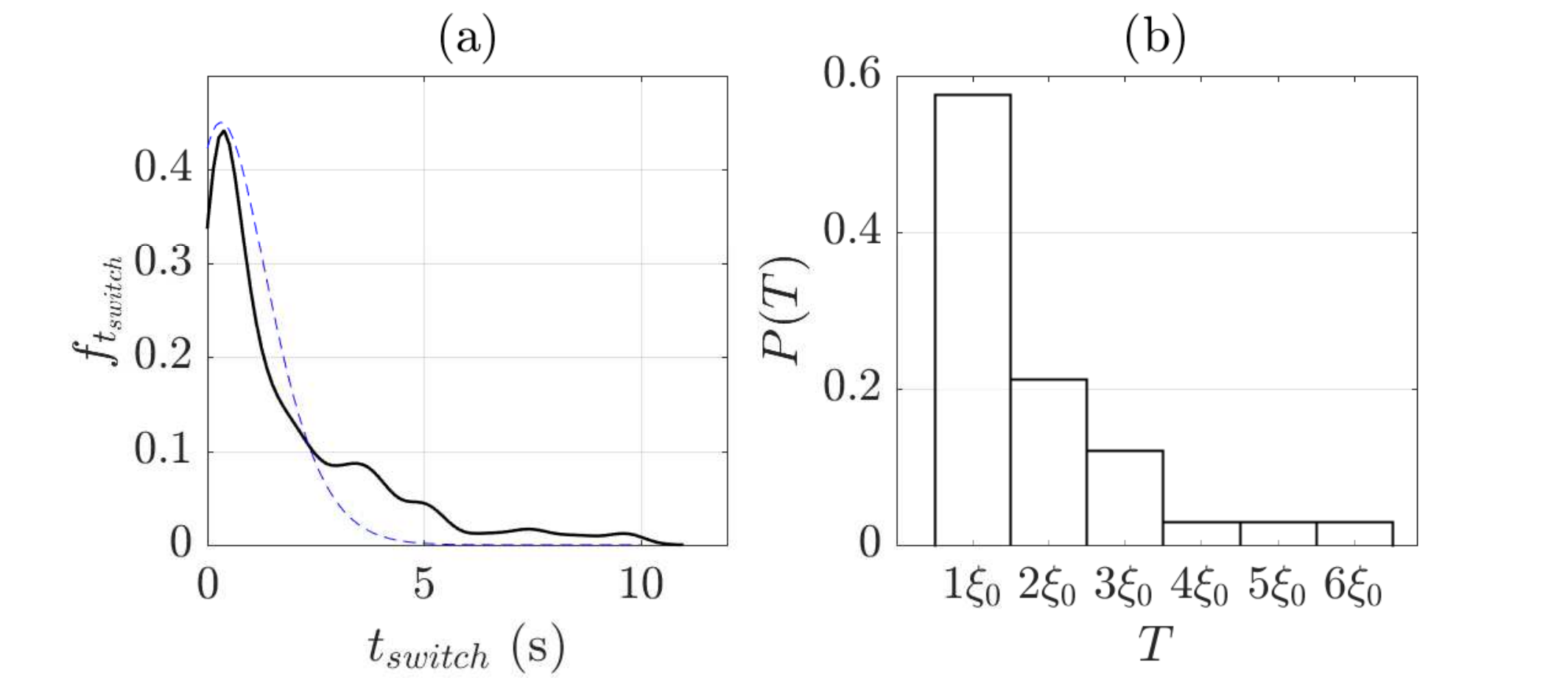}
\caption{(a) PDF of the switching time $t_{switch}$ for the raw data (black solid line) together with a Poisson distribution (parameter set to $0.86$, blue dashed line) and (b) probabilities of switching at times $T$ multiple to $\xi_0=4.3$ for the LP filtered signal.}\label{fig:telegraph_switchTime}
\end{figure}

Furthermore the PDF of the switching times $t_{switch}$ of the raw time series is given in Fig. \ref{fig:telegraph_switchTime} (a), revealing itself close to a Poisson distribution whose parameter is $0.86$ s. The mean switching time, which is the parameter of the Poisson distribution, is of the same order as the residential time $T_{RSB}$ seen previously. Regarding the corresponding LP filtered data, the probability of switching $P(T)$ occurring at times $T$ multiple to $\xi_0$ is computed in Fig. \ref{fig:telegraph_switchTime} (b). The results show that the probability of switching at time $T=\xi_0$ is $P(\xi_0)=0.56$, agreeing with the ACF approximation of Eq. (\ref{eq:linearApproxTelegraphQR}). \citep{Anishchenko2002} obtained $P(\xi_0)=0.52$ for the Lorenz system of Eq. \ref{eq:LorenzDef}, which is also close to $0.5$ as expected for such a strange chaotic attractor.

\section{Conclusions and perspectives}

The dynamics of the near wake behind a squareback bluff body are characterized by the time-evolution of its intensity recirculation barycenter and the spatially-averaged rear pressure barycenter. Both quantities track large-scale structures and exhibit strong bimodal distributions characteristic of a random switching process between two states. Their respective spanwise dynamics are highly anti-correlated (phase opposition). Plotting the trajectories of the pressure barycenter over the rear-part of the model, similar to a Poincar\'e section, stochastic behavior is first observed. But applying the same process to the low frequencies signal reveal a chaotic aspect of the dynamics. The instantaneous pressure barycenter circles around two stable areas acting like strange attractors and randomly switches from one attractor to the other. We characterize the chaotic dynamics of these barycenters by reconstructing their phase space and computing the largest Lyapunov exponent and the correlation dimension. All of these elements tend to describe the dynamics of a complex 3D turbulent wake as a weak chaotic system.
Apart from its fundamental interest, this result is also of great practical interest. Indeed, if the wake dynamics can be modeled as a chaotic attractor, it opens the path to many closed-loop control strategies which have been first tested on simple chaotic systems such as the Lorenz system \cite{Kaiser2014,Guéniat2016}. Recently, a machine learning control based on the low pass filtered signal of the rear pressure of the Ahmed body has been successfully performed \cite{Li2016} and a similar one based on the low frequencies dynamics of the wake can be envisioned \cite{Gautier2015c}.

\section{Acknowledgments}
We deeply acknowledge Diogo Barros, Yacine Bengana, Olivier Cadot, Pierric Joseph and Laurette Tuckerman for their valuable stimulating discussions.

\bibliography{ref_varon_v2}

\end{document}